\documentclass[preprint,prd,aps,nofootinbib]{revtex4}
\usepackage{latexsym}
\usepackage{amssymb}
\usepackage{bm}
\newcommand{\be}{\begin{equation}}
\newcommand{\ee}{\end{equation}}
\newcommand{\bea}{\begin{eqnarray}}
\newcommand{\eea}{\end{eqnarray}}
\newcommand{\bt}{\begin{tabular}}
\newcommand{\et}{\end{tabular}}
\newcommand{\ba}{\begin{array}}
\newcommand{\ea}{\end{array}}

\newcommand{\reals}{{\mathbb{R}}} 

\def\cstok#1{\leavevmode\thinspace\hbox{\vrule\vtop{\vbox{\hrule\kern1pt
\hbox{\vphantom{\tt/}\thinspace{\tt#1}\thinspace}}
\kern1pt\hrule}\vrule}\thinspace}

\begin{document}

\begin{center}
\bibliographystyle{article}
{\Large \textsc{Nonlocal field theory driven by a deformed product.
Generalization of Kalb--Ramond duality}}
\end{center}

\author{Elisabetta Di Grezia$^{1,2}$\thanks{
Electronic address: digrezia@na.infn.it},
Giampiero Esposito$^3$\thanks{
Electronic address: giampiero.esposito@na.infn.it},
Gennaro Miele$^{4,3}$\thanks{
Electronic address: gennaro.miele@na.infn.it}}

\affiliation{
${ }^{1}$Universit\`a Statale di Bergamo, Facolt\`a di Ingegneria,
Viale Marconi 5, 24044 Dalmine (Bergamo), Italy\\
${ }^{2}$Istituto Nazionale di Fisica Nucleare, Sezione di Milano,
Via Celoria 16, 20133 Milano, Italy\\
${ }^{3}$Istituto Nazionale di Fisica Nucleare, Sezione di Napoli,
Complesso Universitario di Monte S. Angelo, Via Cintia
Edificio 6, 80126 Napoli, Italy\\
${ }^{4}$Dipartimento di Scienze Fisiche, Complesso Universitario
di Monte S. Angelo,  Via Cintia Edificio 6, 80126 Napoli, Italy}

\vspace{0.4cm}
\date{\today}

\begin{abstract}
A modification of the standard product used in local field theory
by means of an associative deformed product
is proposed. We present a class of deformed
products, one for every spin $S=0,1/2,1$, that induces
a nonlocal theory, displaying different form for
different fields. This type of deformed product is
naturally supersymmetric and it has an intriguing duality.
\end{abstract}

\maketitle
\bigskip
\vspace{2cm}


\section{Introduction}

The main problem of any local quantum field theory is the presence of
ultra-violet divergences. In fact the S-matrix is expressed in
terms of the products of causal functions of the field
operators. Since the causal functions have fairly strong
singularities on the light cone, the products of such functions
are not mathematically defined. This problem arises from the
ill-defined nature of the product of two local field operators at the
same space-time point. This generates one of the main problems of
quantum field theory - the so-called problem of ultraviolet divergences.
There are different regularization procedures to deal with such
divergences, making the S-matrix elements mathematically
meaningful. These are, for example, the subtraction procedure
interpreted in various manners in a local field theory, the
summation of asymptotic series for the Green functions and
super-propagators, a nonlocal generalization of the theory. In
particular, the nonlocal quantum field theory which replaces
local quantum field theory is very old, dating from 1950's,
starting with Pais and Uhlenbleck (1950), Efimov and coworkers
\cite{efim} (1970-onwards), Moffat, Woodard and coworkers (1990)
\cite{kleppe}, \cite{moffat}. The basic idea to try to avoid
``infinities'' was to assume a nonlocal interaction and thus to
provide a natural cut-off. One has to build a nonlocal quantum
field theory which is a self-consistent scheme satisfying all
principles of conventional quantum field theory (unitarity, causality,
relativistic invariance, etc.) and providing the basis for correct
description of nonlocality effects.
There are some problems with the gauge invariance because of
nonlocality in gauge field interactions. Nevertheless, there are
different nonlocal approaches which give a good possibility of
building a completely ultraviolet finite theory of fundamental
interactions.

(i) One way is to introduce nonlocality in the interaction term
\cite{nonloc0} writing down the Lagrangian of scalar fields in the form
\begin{equation}
L=\phi(x)^\dag(\partial^2+m^2)\phi(x) + \lambda\Phi(x)^\dag\Phi(x),
\label{(1)}
\end{equation}
where the nonlocal field $\Phi(x)$ is obtained from the local
one $\phi(x)$ by ``smearing'' over the nonlocality domain with
the characteristic scale $l_0$. Without specifying the nature of
this nonlocality, and introducing the phenomenological form factor
$K$, the nonlocal field $\Phi(x)$ is defined as
\begin{equation}
\Phi(x) \equiv \int dy K(x-y)\phi(y)=K(l_0^2\partial^2)\phi(x),
\end{equation}
where the nonlocal operator $K(l_0^2\partial^2)$ can be written
in the form
\begin{equation}
K(l_0^2\partial^2)=\sum_{n=0}^{\infty}\frac{c_n}{(2n)!}
(l_0^2\partial^2)^{n},
\end{equation}
$K$ being an entire function without any zeros.
Then the generalized function $K(x-y)=
K(l_0^2\partial^2)\delta(x-y)$ belongs to one of the spaces of
nonlocal generalized functions which was introduced and explored
in the works of Efimov \cite{efim}. One rewrites the Lagrangian
(\ref{(1)}) in terms of the nonlocal fields $\Phi(x)$ as
\begin{equation}
L=\Phi(x)^\dag(\partial^2+m^2)\Phi(x) + \lambda\Phi(x)^\dag\Phi(x),
\label{(2)}
\end{equation}
in this way the $\phi(x)$-propagator (smeared propagator) is
obtained by taking the Fourier transform of
\begin{equation}
\frac{\exp{\left[\frac{p^2-m^2}{l_0^2}\right]}}{p^2-m^2+i\epsilon}.
\end{equation}
This suggests interpreting the nonlocal quantum field theory
as an effective theory valid up to an energy scale $l_0$; and
for energy scales beyond $l_0$, one has to replace
the nonlocal quantum field theory by
a more fundamental theory of constituents having its own larger
mass scale and coupling constant. In a sense, this formulation can
be viewed either as a regularization, or as a physical theory
with a finite mass parameter $l_0$. Such a theory preserves
causality at tree-level in the S-matrix (it is the same as the local
one), but it suffers from quantum causality violations, which are a
serious limitation.

(ii) Another way to have a finite
quantum field theory was proposed in \cite{Nonloc}
on the basis of infinite-component fields, which results in the
introduction of a special form of nonlocality.

(iii) Another way to introduce nonlocality in the theory is to consider
a special class of field theories with higher derivatives. In the
canonical formulation, one usually considers Lagrangians with only
first derivatives. However, higher derivative theories, including
nonlocal theories, also have many physical applications. For
example, when one integrates out high energy degrees of freedom in
a local field theory, the low-energy effective action is
generically nonlocal \cite{[1.1]}. Higher derivative theories were
also considered in order to find a finite quantum field theory
\cite{[2.2]}, before the advent of renormalization. Moreover,
theories with infinitely many derivatives are unavoidable in string
theory \cite{[3.3],[4.4]}. There are other examples, such as
higher derivative gravity \cite{[5.5]}, meson-nucleon interactions
\cite{[6.6]}, and spacetime noncommutative field theory
\cite{[7.7],[8.8]}, and so on. In most cases, higher derivative
terms appear as higher-order corrections in the effective
Lagrangian, hence a perturbative approximation scheme would already
be very useful.

(iv) Another example of nonlocality is in quantum mechanics where
it has long been an intriguing topic in the past decades, and so
far there has been no experiment contradicting
nonlocality. It refers to the correlation between two particles
separated in space, e.g. the entanglement derived from the Bell
theory \cite{bell} and well confirmed in many experiments
\cite{bell1}. All these experiments used massless photons as
carriers of the states, and the nonlocality is of the Bell type.
\\
Finally we would like to mention our approach, in which nonlocality
is introduced through the deformation of the product, and in a
perturbative approximation it is reduced to a field theory with
higher derivatives.

We shall now introduce the plan of the paper.  In Sec. 2 we
review the deformed products, i.e. deformed field theories and
their nonlocal properties. In Sec. 3 we propose a class of
deformed products which are associative, with different expressions
for spin $S=0,1/2,1$, respectively. In Sec. 4 we exhibit
their properties. In Sec. 5 we study the deformed interaction term. In
Secs. 6 and 7 we show that the scalar field theory with our deformed
product is the same as found by Moffat \cite{moffat1}.
Concluding remarks are presented in Sec. 8.

\section{Review of deformed field theory}

Deformation quantization was born as an attempt to interpret the
quantization of a classical system as an associative deformation
(i.e. via star-products) of the algebra of classical observables.
This idea was behind the
mind of many mathematical physicists and physicists
\cite{weyl,wigner} as illustrated by the historical developments
which led to deformation quantization. By deformed it is meant
that the standard point-wise multiplication of functions has been
replaced by a new product which may or may not be commutative.
Recently, algebras of functions with a deformed
product have been studied intensively \cite{cit}. These are
deformed (star-) products which remain associative but not
commutative. There is a class of K-deformed products which
generate deformed associative products, see for example
\cite{pvm}. Before talking about deformed field theory we first
summarize the concept of deformed coordinate spaces as quantum
spaces. Deformed coordinate spaces are defined
in terms of coordinates $x_\mu$ and their commutation
relations. The $\theta$, $k$, $q$-deformations are the best known
examples \cite{dcs}. They are: the canonical relations
$[x_\mu,x_\nu]=\theta_{\mu\nu}$, which for constant
$\theta_{\mu\nu}$ leads to the so-called
$\theta_{\mu\nu}$-deformed coordinate space; the
Lie-type relations where the coordinates form a Lie algebra
$[x_\mu,x_\nu]=C_{\mu\nu}^\lambda x_\lambda$ and
$C_{\mu\nu}^\lambda$ are the structure constants,
this framework leads to the $k$-deformed coordinate space;
and then the quantum group relations,
$x^\mu x^\nu= R_{\rho\sigma}^{\mu\nu}x^\rho
x^\sigma\theta_{\mu\nu}$ where the R-matrix defines a quantum
group, this leads to the q-deformed spaces.

Deformations of mathematical structures have been used at
different moments in physics. When Galilean transformations
between inertial systems were seen not to describe adequately the
physical world, a deformation of the group law arose as the
solution to this paradox. The Lorentz group is a deformation of the
Galilei group in terms of the parameter $c$ . In this deformation
scheme, the old structure is seen as a limit or contraction when
the parameter takes a preferred value. Hence a deformation, an
inverse of contraction (in the sense of Segal--Wigner--Inonu
contraction), is one of the methods of generalization of a
physical theory \cite{[1]}. The undeformed theory can be recovered
from the deformed one when taking a limit of deformation parameter
to some value, e.g., nonrelativistic, classical physics, the
undeformed theory, is recovered from relativistic physics when
taking the velocity of light $c\rightarrow \infty$, and, from
the point of view of quantum physics,
when taking the Planck constant $h \rightarrow 0$.
The mathematical structure of quantum mechanics has also an
ingredient of deformation with respect to classical mechanics.
This naive concept has been applied to field theories on
noncommutative spaces considered as deformations of flat
Euclidean or Minkowski spaces. Since noncommutative geometry
generalizes standard geometry in using a noncommutative algebra
of functions, it is naturally related to the simpler context of
deformation theory. The star product is a product in the space of
formal power series in $\hbar$ whose coefficients are functions on the
phase space. Thus, a product of fields on NC spaces can be expressed as a
deformed product or star-product \cite{[2],[3]} of fields on
commutative spaces \cite{[6], [8]}. The star product can be
seen as a higher-order $f$-dependent differential operator acting
on the function $g$. The noncommutativity is governed by a
parameter such that the commutative case appears in the limit where
this parameter approaches zero. The simplest and most well known
example of $\star$-product is the Moyal--Weyl product, and it
first appeared in quantum mechanics \cite{m}. It was first
introduced by H. Weyl for his quantization procedure and later by
Moyal \cite{[1]} to relate functions on phase space to quantum
mechanical operators in Hilbert space. For this reason the
$\star$-product is called in the literature as Moyal--Weyl product. In
the Moyal--Weyl $\star$-product representation the noncommutative
coordinates ${\hat x}_{\mu}$ (and their functions) are mapped to
commutative coordinates $x_\mu$ with commutative pointwise product
replaced by deformed (nonlocal) $\star$-product defined as
\begin{eqnarray}
&&f\star g(x) \equiv \exp{(i\theta^{\mu\nu}\partial_\mu^y\partial_\nu^z
f(y)g(z))}|_{y=z=x}\nonumber \\
&=&\sum_{n=1}^{\infty}{\left(\frac{i}{2}\right)}^n
\frac{1}{n!}\theta^{\rho_1\sigma_1}\cdot\theta^{\rho_n\sigma_n}
(\partial_{\rho_1}\cdot\partial_{\rho_n}f(x))
(\partial_{\sigma_1}\cdot\partial_{\sigma_n}g(x)).
\end{eqnarray}
This implies the presence of (infinitely many) derivatives in the
action, hence the theory becomes nonlocal, and the noncommutative
quantum field theories are a special case of a nonlocal quantum
field theory. Other $\star$-products will occur
for different orderings. The way
of looking at noncommutative geometry in terms of deformed
products can give different insights. In fact a deformed gauge
theory leads to a theory with a larger symmetry structure, i.e. the
enveloping algebra structure, and it exhibits its nonlocal nature.
Nevertheless, the commutative field theories can be recovered from
their noncommutative counterparts when the
noncommutativity tensor approaches zero:
$\theta_{\mu\nu}\rightarrow 0$. A property of
noncommutative field theories is the
presence of nonlocal interaction terms,
which explicitly breaks Lorentz invariance.
In fact under the integration, the star-product of fields does not
affect the quadratic parts of the Lagrangian, whereas it gives rise to
a nonlocal interaction part.

Hence, Feynman rules in momentum space are
modified with respect to the commutative ones, in fact the
vertices are modified by a phase factor. The deformed
vertices differ from the nondeformed ones by a factor of type
$\cos( 1/2 p_\mu\theta^{\mu\nu}p_\nu)$. When
$\theta_{\mu\nu}\rightarrow 0$, the deformed vertex reduces to the
nondeformed one.

One can give a star-product quantization scheme following
\cite{marm,marm1}, and see that there is a class of star products
($K$-star products) which are obtained via a specific deformation
procedure \cite{marm1}. Much more recently, noncommutative
geometry has entered physics in different contexts.

One context is string theory. In their pioneering
paper, Connes, Douglas and Schwarz \cite{4} introduced
noncommutative spaces (tori) as possible compactification manifolds
of space-time. Non commutative geometry arises as a possible
scenario for short-distance behaviour of physical theories. In the
framework of open string theory \cite{12}, Seiberg and Witten in
\cite{[7.7]} identified limits in which the whole string dynamics,
in presence of a $B$-field, is described by a deformed gauge
theory in terms of a Moyal--Weyl star product on space-time.
The field theory associated to string theory, in the low-energy limit, is
nonlocal, because the fields in the action are multiplied by a
(deformed) star-product. The deformed theories enjoy
renormalization properties as well as UV/IR connection reminiscent
of string theory. Other approaches connecting
deformation theory to theories of gravity have also appeared in
the literature. Among others, there is the deformation
quantization of M-theory \cite{19}, quantum anti-de Sitter
spacetime \cite{20}, q-gravity \cite{21} and gauge theories of
quantum groups \cite{22}.
Another area covered by noncommutativity is supersymmetric
theory. The deformation aspects of supersymmetric field theories
were investigated in \cite{sc,FL,KL,KP}. Analogous to noncommutative
field theories on bosonic spacetime, noncommutative superfield
theories can be formulated in ordinary superspace by multiplying
functions given on it via a $\star$-product which is generated by
some bi-differential operator or Poisson structure $P$. It defines
a deformed superspace and leads to deformed products for general
superfields. There will be symmetries of the undeformed (local)
field theory which are explicitly broken in the deformed (nonlocal)
case. In this case only free actions preserve all supersymmetries
while interactions get deformed and are not invariant under all
standard supersymmetry transformations, because the integral of the star
product of two superfields is not deformed, while in the case of three
or more superfields the integral is deformed.

In \cite{FL} the authors present a variety of deformations, both
for N = 1 and extended (N = 2) supesymmetry in D = 4, which vary
according to the differential operators chosen to construct the
Poisson bracket that afterwards becomes quantized with a star
product of Moyal--Weyl type. For example
in \cite{FL}, the first deformation has the advantage of
being manifestly supersymmetric, while the second, although it
explicitly breaks half of the supersymemtry, allows the
definition of chiral and antichiral superfields, which form
subalgebras of the star product. Another way to
construct a deformed field theory is with derivatives which are an
essential input for the construction of deformed field equations
such as the deformed Klein--Gordon or Dirac equations \cite{gord}.

At the end we would like to emphasize some properties of deformed field
theory as a nonlocal theory. The quantum deformation
modifies the behavior of relativistic theories at distances
comparable to and smaller than the length $l$ corresponding to the
deforming parameter. It appears that, by virtue of the deformation of
local product of fields in the interaction, the vertex will be
replaced by a deformed nonlocal product, with the nonlocality
extending to distances of order $l$.

Such a quantized space-time geometry can provide additional
convergence factors or even a finite quantum field theory. Indeed,
if one introduces a masslike deformation parameter, it occurs also
as a regularizing parameter.

There are also attempts to remove the ultraviolet divergences by
introducing nonlocality into the interaction Lagrangian.
Hence the advantage of the nonlocal character of the deformed
product is the following: first, one has succeeded in
introducing into the interaction Lagrangian all the ambiguity in
the choice of the shape and the value of the ``elementary'' length;
second, the amplitudes of the physical processes have no
additional singularities in the finite region of change of the
invariant momentum variables as compared to the local theory.

Nonlocal quantum field theory faces, however, many difficulties.
One of the main difficulties in constructing the
non-local quantum field theory appears to be the formulation of
macro-causality of the $S$-matrix. Then it seems that a
reasonable macro-causality condition imposed on the $S$-matrix
would be a generalization of the micro-causality condition
\cite{efim}. However, as one can see in Efimov's paper \cite{efim},
there is indeed a causality violation but,
from the physical point of view,
the problem may be formulated in such a way that the amount of
causality violation would satisfy the usual requirements imposed
on nonlocal theories. Indeed, using the Lagrangian of the
quantized field system, Efimov expands the
$S$-matrix in the small coupling constant. Therefore, in the case
of the small coupling constant interaction, the violation of
causality at large distances is rather small.
Another property is the unitarity. The postulate of unitarity of
the $S$-matrix in quantum field theory is one of the principal
requirements for the theory to be regarded as self-consistent and
physically acceptable. Efimov for example, in Ref.
\cite{efim1}, proves the unitarity of the S-matrix in the $n$-th
order of perturbation theory in a nonlocal quantum field theory.

\section{A class of deformed products}

In this section we propose a class of associative deformed products,
with different expressions for spin $S=0,1/2,1$, respectively.
In \cite{marm1} a deformed operator product was
introduced ($K$-product) in the form $\hat{f} ·_K \hat{g} =
\hat{f} \hat{K} \hat{g}$ where $\hat{K}$ is a generic operator. It
satisfies the associativity condition
\begin{equation}
(\hat{f} \bigtriangleup_{K} \hat{g})
\bigtriangleup_{K} \hat{h}=\hat{f}\bigtriangleup_{K}
(\hat{g}\bigtriangleup_{K} \hat{h}).
\end{equation}
As emphasized in \cite{pvm} the K-deformed products are a way to
generate new associative products. To construct our new deformed
product we take into consideration this one and that the
star-product in Quantum Mechanics, because of its nonlocal nature,
can be described through an integral kernel \cite{star}. This
integral kernel plays the role of the structure function for the
product and the star-product reduces to the more familiar
asymptotic expansion with a particular choice of this kernel.
Inspired by all these properties, and bearing in mind that the
star product is a particular associative deformed product, we
import this formalism used in Quantum Mechanics to Field Theory to
define a new deformed product $(A\diamondsuit_\theta B)(x)$
according to
\begin{equation}
(A\diamondsuit_\theta B)(x)\equiv \int_{\reals^4}\int_{\reals^4}\,
A(y)L(x,y,z)\,B(z) \,dy\,dz, \label{Moyal-prodint}
\end{equation}
where the integral kernel $L$ has different shape for each
spin. The associativity condition for operator symbols
implies that the kernel $L(x,y,z)$ satisfies the nonlinear equation
\begin{equation}
\int L(x_1,x_2,y)L(y,x_3,x_4)dy= \int L(x_1,y, x_4)L(x_2,x_3,y)dy.
\label{ass}
\end{equation}
Our kernel is of the type  $\delta(x,z)[\exp{\theta
f(\partial})]\delta(x,y)$, hence it fulfils the associativity
condition (\ref{ass}) thanks to the properties of a Dirac
$\delta$-functional.

For every spin we have a different choice of integral kernel.
In the case of $S=0$, i.e. a scalar field, we start with a theory
with a static term
\begin{equation}
\int d^4 x \frac{1}{2}m_2^{2}.
 \phi\diamondsuit\phi .
\end{equation}
With a particular choice of kernel, we can write
\begin{equation}
\delta L_{{\rm scal}}=\delta(x,z)\left[-1 + \theta\Box\right]\delta(x,y),
\end{equation}
and we obtain a dynamical theory. It can be seen as a leading
order term of expansion, in the parameter $\theta$,
of the following general definition:
\begin{equation}
L_{{\rm scal}}:= \delta(x,z)[\exp{\theta\Box}]\delta(x,y).
\label{freesca}
\end{equation}
In this form it actually displays a ``static nature'', as we will
show later.

In the case of $S=1$, i.e. a vector field, we start with a theory
with a static term
\begin{equation}
 \int
d^4 x \frac{1}{2}m_1^2
 A_\mu \diamondsuit A^{\mu}.
\end{equation}
The particular choice of kernel is now such that
\begin{eqnarray}
&&\delta L_{{\rm
vect}}(x,y,z)=\delta(x,z)[D_\theta]_\mu^\nu\delta(x,y)
=\delta(x,z)\left[g_\mu^\nu +
\theta\Delta_\mu^\nu\right]\delta(x,y)\nonumber
\\
&& =\delta(x,z)\left[g_\mu^\nu + \theta  \left(\Box_y
\delta_\mu^\nu- \left(1-\frac{1}{\alpha}\right)
\nabla^\nu\nabla_\mu\right)\right]\delta(x,y),
\end{eqnarray}
which leads to a dynamical theory. It can be seen as a leading-order
term in the expansion, in the parameter $\theta$, of the following
general definition:
\begin{equation}
L_{{\rm vect}} := \delta(x,z)[\exp{\theta\Delta^\mu_\nu}]\delta(x,y).
\end{equation}

In the case of spin $S=1/2$, i.e. a spinor field, we start with a
theory with a static term
\begin{equation}
 \int
d^4 x \frac{1}{2}m_3 \psi\diamondsuit\overline{\psi}.
\end{equation}
Our particular choice of kernel is such that
\begin{equation}
\delta
L_{{\rm matter}}(x,y,z)=\delta(x,z)\left[
-1+i\sqrt{\theta}\gamma_\mu\partial^\mu\right]\delta(x,y).
\end{equation}
It can be seen as a leading-order term in the
expansion, in the parameter $\theta$, of the following general
definition:
\begin{equation}
L_{{\rm matter}}:=
\delta(x,z)[\exp{-i\sqrt{\theta}\gamma_\mu\partial^\mu}]\delta(x,y).
\label{fermionic}
\end{equation}

\section{Properties of the deformed product}

In this section we analyze some peculiar properties of our
deformed product, outlining the possible implications.

\subsection{Supersymmetric nature of the deformed product and its
apparent dynamical nature}

To achieve correspondence to lowest-order theory we must impose
the condition: $m_i\sqrt{\theta}=1$, which then implies
$m_i=m=\frac{1}{\sqrt{\theta}}$. This requires, of course,
supersymmetry, i.e. that the fields $\phi, \psi, A_\mu$ belong to
a massive vector $N=1$ superfield. Hence, to lowest order, the
``static'' massless theory in the corresponding deformed product
for scalar, vector, fermionic fields is ``equivalent'' to a dynamical
massive supermultiplet, i.e., if we restrict ourselves to the
leading term, linear in the deformation parameter, we get
naturally a supersymmetric formulation, as in a dual deformation.

In principle one can think of having a class of deformed products
and, corresponding to different choices of kernel, to build a
deformed supersymmetryc Wess--Zumino model,
as in a supersymmetric U(1) theory,
introducing the dynamics to lowest order in the kernel
and not in the supersymmetric formulation, as in \cite{sup}. In
our case, with suitable choice of kernel in the scalar, spinor
and vector Lagrangian,
we can reproduce in a natural way a supersymmetric
action which suffers from nonlocality at subsequent orders in
$\theta$. The Lagrangian for a globally supersymmetric matter
multiplet is \cite{mukhi}
\begin{eqnarray}
L^{N=1}_{{\rm matter}}=-\frac{1}{4}F_{\mu\nu}F^{\mu\nu} +
\frac{1}{2}\bar{\psi}\gamma_\mu\partial^\mu\psi -
\frac{1}{2}m\bar{\psi}\psi + \frac{1}{2}m^2A_\mu A^\mu -
\frac{1}{2}m^2\phi^{2},
\end{eqnarray}
where $A^{\mu}$ is a vector field, $\psi$ is a Majorana spinor
field, and $\phi$ is a pseudoscalar field.

All fields must have the same mass. The above Lagrangian in
superspace formalism is
\begin{equation}
L^{N=1}_{{\rm matter}}= [\Phi\dag \Phi]_D + m[\Phi \Phi]_F
+m[V_{WZ}^2]_D\nonumber + \frac{1}{32}[W^\alpha W_\alpha]_{F}.
\end{equation}
With our prescription, it becomes
\begin{equation}
L^{N=1}_{{\rm matter}}= \frac{1}{2}m^2
A_\mu \diamondsuit_{S=1} A^\mu
+ \frac{1}{2}m \psi\diamondsuit_{S=1/2}\overline{\psi }+\frac{1}{2}m^2
\phi\diamondsuit_{S=0}\phi.
\end{equation}
What seems to happen is that the dynamics, which is put at the
level of superfields in a supersymmetric Lagrangian, is found in a
nonlocal theory to first order non vanishing in the $\theta$
parameter.

Hence one can think of using the deformed product, to lowest order
in the $\theta$ parameter, to obtain the dynamics, i.e.,
one can encode dynamics in a product and the other way around.
Changing deformed product means having a different dynamics.
We can have infinitely many derivatives in order
to reproduce, at different orders in $\theta$, the dynamics of the
system. This is a first step towards obtaining a more elaborate
model and the dynamics in the deformed product, where the
$\theta$ parameter is essential, and its smallness is important to
make sure that higher-order (derivative) terms are of no importance.
Thus, {\it the dynamics can be seen as a ``perturbative'' effect
which disappears in the global ``nonperturbative'' static
expression}.

\subsection{Nonlocal nature of the deformed product}

Hence, starting with the free static action and using the
(bi)-differential operator to lowest order in the parameter, one
obtains a known massive free-field theory,
that is a lower order approximation
of a more complicated nonlocal theory. In this way we have a
nonlocal field theory from a local theory in which we deformed
the product on a nonlocality domain with a characteristic scale
$\sqrt{\theta}$. While in \cite{kovalenko}, \cite{moffat},
\cite{prokhorov}, \cite{joglekar} one replaces a local field by a
``smearing'' field, in our model we put the nonlocality in the
product, as in the star product. Thus, the deformed quantum field theory
is a particular case of a nonlocal quantum field theory.

\subsection{Duality of the Kalb--Ramond field in this deformed product}

In this section we analyze an intriguing
property of a type of deformed product. We show how the duality
property of a Kalb--Ramond is changed by using a particular deformed
product at lower order in $\theta$. A well-known result is that a
massless field $H_{\nu\rho\sigma}=\partial_{[\nu}B_{\rho\sigma]}$
in an undeformed product is equivalent to a massless scalar field,
i.e., the degrees of freedom of the antisymmetric tensor field
$B_{\rho\sigma}$ are only one:
\begin{equation}
\partial^{\mu}\phi=\frac{1}{6}\epsilon^{\mu\nu\rho\sigma}H_{\nu\rho\sigma},
\end{equation}
while the massive field is equivalent to a massive vector field,
i.e., the degrees of freedom of $B_{\rho\sigma}$ are three:
\begin{equation}
H^{\mu}=\frac{1}{6}\epsilon^{\mu\nu\rho\sigma}H_{\nu\rho\sigma}.
\end{equation}
In view of these properties, we want to generalize the duality of
Kalb--Ramond in the deformed case.
We show that a massless deformed Kalb--Ramond
theory is dual to a U(1)-breaking theory, i.e., it is equivalent
to a massive vector field theory, with the above choice of
duality and choosing a particular kernel for the deformed product.
The presence of deformation leaves nontrivial transverse and
longitudinal modes, unlike the classical undeformed massless case.
For this purpose, we are interested in a model ruled by the
deformed action
\begin{equation}
S_{H}=\frac{1}{3!}\int d^{4}x H \star_\theta H(x) := \int d^{4}x
\left[\int_{\reals^4}\int_{\reals^4}\,
H_{\alpha\beta\gamma}(y)
L^{\alpha\beta\gamma}_{\rho\sigma\tau}(x,y,z)\,H^{\rho\sigma\tau}(z)
\,dy\,dz\right],
\label{Moyal-prodint}
\end{equation}
while the action $S_{B}=\frac{1}{3!}\int
d^{4}xH_{\mu\nu\rho}(B)H^{\mu\nu\rho}(B)$ is deformed through the
integral kernel $L$ chosen to recover
the U(1) gauge theory, and is given by
\begin{equation}
L^{\alpha\beta\gamma}_{\rho\sigma\tau}(x,y,z):=
\frac{1}{6}\epsilon^{\mu\alpha\beta\gamma}
\epsilon_{\nu\rho\sigma\tau}[D_\theta]_\mu^\nu,
\label{Moyalkernel}
\end{equation}
and $[D_\theta]_\mu^\nu$ reads as
\begin{eqnarray}
[D_\theta]_\mu^\nu
& := & \delta(x,z)\left[g_\mu^\nu +
\frac{\theta^2}{m^2}\Delta_\mu^\nu\right]\delta(x,y) \nonumber \\
& = & \delta(x,z)\left[g_\mu^\nu
+ {\theta^{2}\over m^{2}} \left(\Box_y \delta_\mu^\nu -
\left(1-\frac{1}{\alpha}\right)
\nabla^\nu\nabla_\mu\right)\right]\delta(x,y).
\label{Moyalkernel21}
\end{eqnarray}
The presence of $\alpha$ in (\ref{Moyalkernel21}) reflects what we
know about QED in its formulation with functional integrals in the
Lorenz gauge to obtain an invertible operator on the potential
$A_\mu$. Actually, its inclusion for a massive vector field model
is not compelling, it is a redundant term, i.e., we are summing a
vanishing term ($\alpha=\infty$). With this choice of $L$ and with
the following duality transformation:
\begin{equation}
H_{\mu\nu\rho}=\frac{1}{\sqrt{\theta}}\epsilon_{\mu\nu\rho\sigma}A^{\sigma},
\label{dual}
\end{equation}
to first order in $\theta$, the fundamental (at high energy)
deformed action (\ref{Moyal-prodint}) is dual to the action of a
massive spin-$1$ field with mass $m_\theta
=\frac{1}{\sqrt{\theta}}\sim M_{pl}$ (effective theory at low
energy), i.e.
\begin{equation}
S_{\rm dual}=S_H=\int d^{4}x\left[
\frac{1}{2}m_{\theta}^2 A_\mu A^\mu
-\frac{1}{4}F_{\mu\nu}F^{\mu\nu} + \frac{1}{2\alpha}(\partial_\mu
A^\mu)^2\right],
\label{azioneqed}
\end{equation}
where $m=\frac{1}{\sqrt{\theta}}$ and
$\partial_\mu A^\mu$ can be shown to vanish.

In general we can give a formal expression of $L$ to every order
in $\theta$:
\begin{equation}
[D_{\rm global}]_\mu^\nu:= \delta(x,z)[\exp{{\theta^{2}\over
m^{2}}\Delta_\mu^\nu}]\delta(x,y). \label{Moyalkernelglobal1}
\end{equation}
Its expansion in powers of $\theta$ is in terms of derivatives of
increasing order.

To zeroth order in $\theta$ (or $\theta=0$), we recover the
duality of the massless Kalb--Ramond to a massless scalar field
$\phi$, putting  $A^{\sigma}=\partial^\sigma \phi$. To first order
in ${\theta^{2}\over m^{2}}=\tilde{\theta}$, a deformed massless
Kalb--Ramond (gauge invariant) is dual to a massive spin-$1$
field.

The operator $\Delta_\mu^\nu$ in Eq. (\ref{Moyalkernelglobal1}) is
a hyperbolic operator. To have a meaningful expression, we have to
make a Wick rotation, so that it becomes
an elliptic operator.
In this way the action of the massive vector field can be seen as
the lowest order (effective theory) of a (broader) nonlocal
theory, which is reduced at zeroth order to a free scalar theory.
A massive spin-$1$ theory may be regarded as the low-energy limit
of a fundamental deformed theory, where the low-energy limit is
set by the massive term $\frac{1}{\sqrt{\theta}}$.

On using different representations of deformed product it is possible to
find a ``dual'' deformed product, i.e., we can use duality in the reverse
order. Then we start with $\int d^4 x \frac{1}{2}m_1^2
A_\mu \diamondsuit A^\mu$, and after the duality transformation we
have $\int d^4 x \frac{1}{2}H
\diamondsuit H$, hence we are able to write a dual deformed product
to first perturbative order. In this way the duality and the
deformation are connected, and when $\theta\rightarrow 0$, the
deformed product and the duality disappear.

\section{Deformed interaction term}

In this section we analyze the deformed interaction term between
matter and a vector field, and in its dual version.

\subsection{Deformed interaction term between matter and vector field}

We have a U(1) theory of an unknown particle at high energy which
can decay (because we do not observe it) and in principle it can
produce an observable physics. Thus, we can imagine a coupling
with matter and we study the decay. We construct a model in which
the deformed (high energy) action of matter and its
interaction term with the
vector field corresponds to a low-energy action of massive
fermions plus the vector-matter interaction plus correction
derivative terms. The general Lagrangian density for a vector
field $L(V_{\mu },\psi)$ describing all their interactions is given by
\begin{eqnarray}
L &=&-\frac{1}{4}G_{\mu \nu }G^{\mu \nu }
+\frac{1}{2}M^{2}V_{\mu }V^{\mu }
+\beta\partial _{\nu }V_{\mu} V^{\mu }V^{\nu }  \nonumber \\
&&+\gamma V_{\mu }V_{\nu }V^{\mu }V^{\nu }+i \overline{\psi}
\gamma_\mu \partial^\mu \psi -\overline{\psi }m\psi +V_{\mu}
\overline{\psi }\gamma ^{\mu }\psi.
\label{LN1}
\end{eqnarray}
To obtain the correspondence with a vector theory we can consider
different combinations, i.e.,
$(\bar{\psi}\diamondsuit\gamma^{\mu}\psi) A_{\mu}$, or
$(\bar{\psi}\gamma^{\mu}\diamondsuit\psi) A_{\mu}$. We can take
account of two possibilities in the form
\begin{eqnarray}
S_{\psi\bar{\psi}A}&=&\frac{1}{2}\int d^{4}x
\,\left[\bar{\psi}\diamondsuit\left(m -\gamma^{\mu} A_{\mu}
\right)\psi + \bar{\psi}\left(m -\gamma^{\mu}
A_{\mu}\right)\diamondsuit\psi\right] \nonumber \\
& :=& \int d^{4}x \,\left\{\int_{\reals^4}\int_{\reals^4}\,
\left[\bar{\psi}(y) L_{{\rm matter}}(x,y,z)\left(m
-\gamma^{\mu}A_{\mu}\right)\,\psi(z)
\right.\right. \nonumber \\
&&+\bar{\psi}(y) \left.\left.\left(m
-\gamma^{\mu}A_{\mu}\right)L_{{\rm matter}}(x,y,z)\,\psi(z)
\,dy\,dz\right]\right\}, \label{Moyal-prodmatter1}
\end{eqnarray}
hence with the integral kernel $L_{{\rm matter}}$ chosen to recover the
vector-fermion action
\begin{equation}
L_{{\rm matter}}(x,y,z):= \delta(x,z)(-1
+\sqrt{\theta}\gamma_\mu\partial^\mu)\delta(x,y),
\label{Moyalkernel1}
\end{equation}
to first order in $\theta$, the fundamental deformed action
(\ref{Moyal-prodmatter1}) is an action of a massive spin-$1$ in
interaction with matter plus a correction term, i.e.
\begin{eqnarray}
&& S_{\rm}^{{\rm total}}
=S_A + S_{\psi\bar{\psi}A}\nonumber \\ &=&\int d^{4}x\left[
\frac{1}{2}m_{\theta}^2 A_\mu A^\mu
-\frac{1}{4}F_{\mu\nu}F^{\mu\nu}+ i\,m\,\sqrt{\theta}
\overline{\psi }\gamma_\mu
\partial^\mu \psi -\overline{\psi }m\psi\right.
\nonumber \\ &&-\left. \frac{1}{\sqrt{\theta}\,M}A_{\mu }
\overline{\psi }\gamma ^{\mu }\psi - \frac{i}{M} \overline{\psi }
\partial^\rho(A_{\rho }\psi)\right].
\label{azionematter1}
\end{eqnarray}
To have a correct correspondence we have the condition
$m\sqrt{\theta}=1$, i.e., at high energy the only possible mass is
''driven'' by the $\theta$ term, and the charge of coupling is
$Q=\frac{1}{M\sqrt{\theta}}$.

In general we can give a formal expression of $L$ to every order
in $\theta$ as in (\ref{fermionic}):
\begin{equation}
[L_{{\rm matter-global}}]:=
\delta(x,z)[\exp{-i\sqrt{\theta}\gamma_\mu\partial^\mu}]\delta(x,y).
\label{Moyalkernelglobal}
\end{equation}
We thus find that the vector field combines with
matter to become, at low energy,
a massive vector field, and it interacts through the standard
coupling described by the previous Eq. (\ref{azionematter1}). The
presence of deformation ``seems'' to introduce a {\em dynamics} in
a {\em static} system to zeroth order.

\subsection{Deformed interaction term between matter and
Kalb--Ramond field}

We construct a model in which we show that the deformed (high
energy) action of matter and its interaction term with
Kalb--Ramond corresponds, after a duality transformation, to a
low-energy action of massive fermions plus the vector-matter
interaction plus correction derivative terms. Such a tensor field,
which appears, for example, in the massless sector of a heterotic
string theory, is assumed to coexist with gravity in the bulk, in
a five-dimensional Randall--Sundrum scenario \cite{rs}. It has a
well-known geometric interpretation as the spacetime torsion. We
consider the most general gauge-invariant action of a second-rank
antisymmetric Kalb--Ramond tensor gauge theory, including the
coupling with matter modes \cite{rs}:
\begin{equation}
{\cal L}_{\psi\bar{\psi}H}= -\frac{1}{M_{Pl}}
\bar{\psi}[i\gamma^{\mu}\sigma^{\nu\lambda}
H_{\mu\nu\lambda}]\psi.
\end{equation}
The general Lagrangian density for a vector field $L(V_{\mu},\psi)$
describing all their interactions is given by
\begin{eqnarray}
L &=&-\frac{1}{4}G_{\mu \nu }G^{\mu \nu }
+\frac{1}{2}M^{2}V_{\mu }V^{\mu }
+\beta\partial _{\nu }V_{\mu} V^{\mu }V^{\nu }  \nonumber \\
&&+\gamma V_{\mu }V_{\nu }V^{\mu }V^{\nu }+i \overline{\psi
}\gamma_\mu \partial^\mu \psi -\overline{\psi }m\psi +V_{\mu }
\overline{\psi }\gamma ^{\mu }\psi.
\label{LN}
\end{eqnarray}

To obtain the correspondence with a vector theory we can have
different combinations, i.e.,
$\frac{1}{M_{Pl}}(\bar{\psi}\diamondsuit\gamma^{\mu}
\sigma^{\nu\lambda}\gamma_5\psi)
H_{\mu\nu\lambda}$, or
$\frac{1}{M_{Pl}}(\bar{\psi}\gamma^{\mu}\sigma^{\nu\lambda}
\gamma_5\diamondsuit\psi)
H_{\mu\nu\lambda}$. We can take account of two possibilities by writing
\begin{eqnarray}
S_{\psi\bar{\psi}H}^{{\rm dual}}&=&\frac{1}{2}\int d^{4}x
\,\left[\bar{\psi}\diamondsuit\left(m
-\frac{1}{M_{Pl}}\gamma^{\mu}\sigma^{\nu\lambda}\gamma_5
H_{\mu\nu\lambda} \right)\psi + \bar{\psi}\left(m
-\frac{1}{M_{Pl}}\gamma^{\mu}\sigma^{\nu\lambda}
H_{\mu\nu\lambda}\right)\diamondsuit\psi\right] \nonumber \\
& :=&
\int d^{4}x \,\left\{\int_{\reals^4}\int_{\reals^4}\,
\left[\bar{\psi}(y) L_{{\rm matter}}(x,y,z)\left(m
-\frac{1}{M_{Pl}}\gamma^{\mu}\sigma^{\nu\lambda}
\gamma_5H_{\mu\nu\lambda}\right)\,\psi(z)
\right.\right.\nonumber \\
&&+\bar{\psi}(y) \left.\left.\left(m
-\frac{1}{M_{Pl}}\gamma^{\mu}\sigma^{\nu\lambda}
H_{\mu\nu\lambda}\right)L_{{\rm matter}}(x,y,z)\,\psi(z)
\,dy\,dz\right]\right\},
\label{Moyal-prodmatter}
\end{eqnarray}
hence with the integral kernel $L_{{\rm matter}}$ chosen to recover the
vector-fermion action
\begin{equation}
L_{{\rm matter}}(x,y,z):= \delta(x,z)(-1
+\sqrt{\theta}\gamma_\mu\partial^\mu)\delta(x,y),
\label{Moyalkernel}
\end{equation}
and with the duality transformation (\ref{dual})
to first order in $\theta$, the fundamental noncommutative action
(\ref{Moyal-prodmatter}) is dual to the action of a massive
spin-$1$ pseudo-vector field in interaction with matter plus a
correction term, i.e.
\begin{eqnarray}
&& S_{\rm dual}^{{\rm total}}=S_H
+ S_{\psi\bar{\psi}H}^{{\rm dual}}\nonumber \\ &=&\int d^{4}x\left[
\frac{1}{2}m_{\theta}^2 A_\mu A^\mu
-\frac{1}{4}F_{\mu\nu}F^{\mu\nu}+ i\,m\,\sqrt{\theta}
\overline{\psi }\gamma_\mu
\partial^\mu \psi -\overline{\psi }m\psi\right.
\nonumber \\ &&-\left. \frac{1}{\sqrt{\theta}\,M_{Pl}}A_{\mu }
\overline{\psi }\gamma ^{\mu }\psi - \frac{i}{M_{Pl}}
\overline{\psi }
\partial^\rho(A_{\rho }\psi)\right],
\label{azionematter}
\end{eqnarray}
where we have used the identity
$\gamma_\lambda\Sigma_{\mu\nu}=i[g_{\lambda\mu}\gamma_\nu-
g_{\lambda\nu}\gamma_\mu+ i
\epsilon_{\lambda\mu\nu\rho}\gamma_5\gamma_\rho]$. To have a
correct correspondence we have the condition $m\sqrt{\theta}=1$,
i.e., at high energy the only possible mass is ``driven'' by the
$\theta$ term, and the charge of coupling is
$Q=\frac{1}{M_{Pl}\sqrt{\theta}}$.

In general we can give a formal expression of $L$ to every order
in $\theta$:
\begin{equation}
[L_{{\rm matter-global}}]:=
\delta(x,z)[\exp{-i\sqrt{\theta}\gamma_\mu\partial^\mu}]\delta(x,y).
\label{Moyalkernelglobal}
\end{equation}
We thus find that the Kalb--Ramond field combines with
matter to become at low energy a massive vector filed,
and it interacts through the standard
coupling described by the previous Eq. (\ref{azionematter}).

\section{Application of the deformed product to a free scalar field}

In this section we evaluate the Green function and dispersion
relation for the free scalar action
with Lagrangian density (\ref{freesca}), showing its
static nature. Then we analyze the dynamical scalar field theory
with our deformed product, showing that, in a sense, it is equivalent
to the one found by Moffat \cite{moffat1}.

\subsection{Green function and dispersion relation of the nonlocal model.
A fictitious dynamical theory}

Given any differential operator $D$ on $R^4$ one can define a map
$\sigma(D)$ called the {\em symbol} of $D$:
$\sigma:D\rightarrow \sigma(D)\equiv e^{-\alpha k_\mu x^\mu }D
e^{\alpha k_\mu x^\mu }$. The associated equation of motion is
$D\psi =0$, to which there corresponds the dispersion relation
$\sigma(D;k,\omega)=0$, e.g. $\omega=\omega(k)$. For $D=\Box$ one
obtains
\begin{equation}
\sigma(\Box+m^2)=\alpha^2(\vec{k}\cdot\vec{k}-\omega^2) +m^{2}.
\end{equation}
By setting $\sigma(D;k,\omega)=0$ one obtains the dispersion relation
\begin{equation}
E^2=\vec{k}\cdot\vec{k} +m^{2},
\end{equation}
which leads to the following wave equation:
\begin{equation}
(\Box+m^2)\phi=0.
\end{equation}
The Green function is defined by
\begin{equation}
G(x,x')=\int d^4k
\frac{e^{ik_\mu(x^\mu-x'^\mu)}}{\sigma(\Box+m^2)},
\end{equation}
and it satisfies the equation
\begin{equation}
(\Box+m^2)G(x,x')=-\delta(x-x')=-\int d^4{k}
e^{ik_\mu(x^\mu-x'^\mu)}.
\end{equation}
In our nonlocal model for a free scalar field, the symbol is
$\sigma(m^2 \exp{\theta\Box};k,\omega)=m^2\exp{(-k_\mu
k^\mu\theta)}=m^2\sum_{n=0}^\infty (-k_\mu k^\mu\theta)^{n}$. If
we instead consider a finite approximation, at small $\theta$ we have
correction terms. Taking into account that the symbol maps
$\exp{\theta\Box}\rightarrow\exp{-k_\mu k^\mu\theta}$,
the Green function is
\begin{eqnarray}
&&G(x,x')=m^2\int d^4k
\frac{1}{(2\pi)^4}e^{ik_\mu(x^\mu-x'^\mu)}{\exp{(-k_\mu
k^\mu\theta)}} \nonumber \\
&=&  m^2\int d^4k\frac{1}{(2\pi)^4}
e^{\delta^{\mu\nu}(\sqrt{\theta}k_\mu-\frac{i}{2\sqrt{\theta}}(x_\mu-x'_\mu))
(\sqrt{\theta}k_\nu-\frac{i}{2\sqrt{\theta}}(x_\nu-x'_\nu))}
\exp{[-\frac{1}{4\theta}\delta^{\mu\nu}(x_\mu-x'_\mu)(x_\nu-x'_\nu)]}
\nonumber \\
&=& m^2\prod_{m=1}^{4}\int dk_{m}
\frac{1}{(2\pi)^4}e^{\delta^{jl}(\sqrt{\theta}k_{j}
-\frac{i}{2\sqrt{\theta}}(x_{j}-x'_{j}))
(\sqrt{\theta}k_{l}-\frac{i}{2\sqrt{\theta}}(x_{l}-x'_{l}))}
\exp{[-\frac{1}{4\theta}\delta^{\mu\nu}(x_\mu-x'_\mu)(x_\nu-x'_\nu)]}
\nonumber \\
&=& m^2
{\left(\frac{1}{64\pi\theta}\right)}^{2}
\exp{[-\frac{1}{4\theta}
\delta^{\mu\nu}(x_\mu-x'_\mu)(x_\nu-x'_\nu)]},
\end{eqnarray}
where we have used the Gaussian integral, and $G(k)= (m^2\exp{-k_\mu
k^\mu\theta})^{-1}=\frac{1}{-m^2 +k^2 + \sum_{n=2}^\infty (-k_\mu
k^\mu\theta)^n}$, ($\,\,\theta=m^{-2}$), while in \cite{moffat1}
the modified Feynman propagator in momentum space is
$i\Delta_F(k)=\frac{i\exp{1/2k_\mu \tau^{\mu\nu}k_\nu\theta}}{ k^2
 -m^2+i\epsilon}$.
We note that the exact expression of the Green function does not show
any pole, corresponding to the static nature of the global theory,
while dynamics can be recovered from a perturbative expansion.

\section{Application of the deformed product to a free scalar
field theory. Analogy with the Moffat model}

In this section we show that the free scalar field theory with our
deformed product, in a sense, is the same as
that found by Moffat \cite{moffat1}.
As we saw in the above subsection the product does not induce a
dynamics, because the dispersion relation obtained by setting to
zero the symbol does not have a solution.
This means that we have to start with an ordinary dynamical action
rewritten through this deformed product.

Our deformed product can be seen as an application of the product
proposed by Moffat \cite{moffat1} in which we find
supersymmetry in the different shape of the deformed product, but
not in the superspace formalism:
\begin{equation}
\label{diamondproduct}
({\hat\phi}_1\diamondsuit{\hat\phi}_2)(\rho)=\biggl[\exp
\biggl(-\frac{1}{2}\tau^{\mu\nu}
\frac{\partial}{\partial\rho^\mu}\frac{\partial}{\partial\eta^\nu}\biggr)
\phi_1(\rho)\phi_2(\eta)\biggr]_{\rho=\eta} $$ $$
=\phi_1(\rho)\phi_2(\rho)-\frac{1}{2}
\tau^{\mu\nu}\frac{\partial}{\partial\rho^\mu}
\phi_1(\rho)\frac{\partial}{\partial\rho^\nu}\phi_2(\rho)+O(\tau^2).
\end{equation}
Here, by comparison with the Moffat product in \cite{moffat1},
$\tau^{\mu\nu}=\delta^{\mu\nu}\theta$ and $\rho=\eta$. Hence our
product is a particular application of it and it is associative
and commutative. In our case the modified Feynman propagator
${\bar\Delta}_F$ is defined by the vacuum expectation value of the
time-ordered $\star$-product
\begin{equation}
i{\bar\Delta}_F(x-y)\equiv\langle 0\vert
T(\phi(x)\diamondsuit\phi(y))\vert 0 \rangle
=\frac{i}{(2\pi)^4}\int\frac{d^4k\exp[-ik(x-y)]
\exp[\frac{1}{2}(k^2\theta)]}{k^2-m^2+i\epsilon}.
\end{equation}
In momentum space this gives
\begin{equation}
i{\bar\Delta}_F(k)=\frac{i\exp[\frac{1}{2}(k^2\theta)]}
{k^2-m^2+i\epsilon},
\end{equation}
which reduces to the standard commutative field theory form for
the Feynman propagator
\begin{equation}
i\Delta_F(k)=\frac{i}{k^2-m^2+i\epsilon}
\end{equation}
in the limit $\vert\theta^{\mu\nu}\vert\rightarrow 0$. The free-field
$\phi^2$ theory is nonlocal, unlike the corresponding one
in ordinary local field theory, resulting in a modified Feynman
propagator ${\bar\Delta}_F(k)$ and modified dispersion relation.
It turns out to be a particular case of that treated by Moffat in
\cite{moffat1}.

\section{Concluding remarks}

In our paper,
a modification of the standard product used in local field theory
by means of an associative deformed product
has been proposed. We have built a class of deformed
products, one for every spin $S=0,1/2,1$, that induces
a nonlocal theory, displaying different form for
different fields. This type of deformed product is
naturally supersymmetric and it has an intriguing duality.

It now remains to be seen whether a suitable variant of our construction
can lead to a product different from the one used by Moffat
\cite{moffat1}.

\acknowledgments
G. Esposito is grateful to the Dipartimento di Scienze Fisiche of
Federico II University, Naples, for hospitality and support.
Conversations with Fedele Lizzi have been very helpful.


\begin{thebibliography}{}

\bibitem{efim}
G. V. Efimov, {\it Commun. Math. Phys.} {\bf 5} (1967) 42;
Nonlocal Interactions of Quantized Fields (Moscow, 1977).
\bibitem{kleppe}
G. Kleppe and R. P. Woodard, {\it Nucl. Phys. B}
{\bf 388} (1992) 81.
\bibitem{moffat}
J. W. Moffat, {\it Phys. Rev. D} {\bf 41} (1990) 1177;
B. J. Hand and J. W. Moffat, {\it Phys. Rev. D} {\bf 43} (1991) 1896;
D. Evens, J. W. Moffat, G. Kleppe and R. P. Woodard,
{\it Phys. Rev. D} {\bf 43} (1991) 499.
\bibitem{nonloc0}
J. W. Moffat, {\it Phys. Rev. D} {\bf 37} (1990) 1177.
\bibitem{Nonloc}
N. V. Krasnikov, {\it Phys. Lett. B} {\bf 195} (1987) 377.
\bibitem{[1.1]}
A. O. Barvinsky and G. A. Vilkovisky,
{\it Nucl. Phys. B} {\bf 282} (1987) 163;
{\it Nucl. Phys. B} {\bf 333} (1990) 471.
\bibitem{[2.2]}
A. Pais and G. E. Uhlenbeck, {\it Phys. Rev.}
{\bf 79} (1950) 145.
\bibitem{[3.3]}
D. A. Eliezer and R. P. Woodard, {\it Nucl. Phys. B} {\bf 325} (1989) 389.
\bibitem{[4.4]}
H. Hata, {\it Phys. Lett. B} {\bf 217} (1989) 438;
{\it Nucl. Phys. B} {\bf 329} (1990) 698.
\bibitem{[5.5]}
K. S. Stelle, {\it Phys. Rev. D} {\bf 16} (1977) 953;
J. Julve and M. Tonin, {\it Nuovo Cim. B} {\bf 46} (1978) 137.
\bibitem{[6.6]}
P. Kristensen and C. Moller,
{\it K. Dan. Vidensk. Selsk. Mat-Fys. Medd.} {\bf 27} (1952) 7.
\bibitem{[7.7]}
N. Seiberg and E. Witten, {\it J. High Energy Phys.} 99{\bf 09}: 032
(1999).
\bibitem{[8.8]}
N. Seiberg, L. Susskind and N. Toumbas, {\it J. High Energy Phys.}
00{\bf 06}: 044 (2000).
\bibitem{bell}
J. S. Bell, {\it Physics (N.Y.)} {\bf 1} (1964) 195.
\bibitem{bell1}
A. Aspect, P. Grangier and G. Roger,
{\it Phys. Rev. Lett} {\bf 47} (1981) 460;
ibid. {\bf 49} (1982) 91;
P.G. Kwiat, K. Mattle, H. Weinfurter and A. Zeilinger, ibid. {\bf 75}
(1995) 4337;
W. Tittel, J. Brendel, H. Zbinden and N. Gisin, ibid. {\bf 81}
(1998) 3563;
G. Weihs, T. Jennewein, C. Simon, H. Weinfurter and A.
Zeilinger, ibid. {\bf 81} (1998) 5039;
A. Aspect, Nature (London) {\bf 398} (1999) 189.
\bibitem{moffat1}
J. W. Moffat, {\it Phys. Lett. B} {\bf 506} (2001) 193.
\bibitem{weyl}
H. Weyl, {\it The Theory of Groups and Quantum Mechanics}
(Dover, New York, 1931).
\bibitem{wigner}
E. Wigner, {\it Phys. Rev.} {\bf 40} (1932) 749.
\bibitem{cit}
F. Bayen, M. Flato, C. Fronsdal, A. Lichnerowicz and D. Sternheimer,
{\it Ann. Phys.} {\bf 111} (1978) 61;
D. Sternheimer,  {\it AIP Conf. Proc.} {\bf 453} (1998) 107;
M. Kontsevich, {\it Lett. Math. Phys.} {\bf 66} (2003) 157.
\bibitem{pvm}
O. V. Man'ko, V. I. Man'ko, G. Marmo and P. Vitale,
{\it Phys. Lett. A} {\bf 360} (2007) 522.
\bibitem{dcs}
C. S. Chu and P. M. Ho,
{\it Nucl. Phys. B} {\bf 550} (1999) 151;
V. Schomerus, {\it J. High Energy Phys.}
99{\bf 06}: 030 (1999);
J. Lukierski, A. Nowicki, H. Ruegg and V. N. Tolstoy,
{\it Phys. Lett. B} {\bf 264} (1991) 331;
J. Lukierski, A. Nowicki and H. Ruegg,
{\it Phys. Lett. B} {\bf 293} (1992) 344;
B. Jurco, S. Schraml, P. Schupp and J. Wess,
{\it Eur. Phys. J. C} {\bf 17} (2000) 521;
A. Klimyk and K. Schmudgen, {\it Quantum Groups and
Their Representations} (Springer, Berlin, 1997).
\bibitem{[1]}
J. E. Moyal, {\it Proc. Cambridge Phil. Soc.} {\bf 45} (1949) 99.
\bibitem{[2]}
F. Bayen, M. Flato, C. Fronsdal, A. Lichnerowicz and D.
Sternheimer, {\it Ann. Phys. (N.Y.)} {\bf 110} (1978) 111.
\bibitem{[3]}
C. Zachos, Geometrical evaluation of star
products, hep-th/9912238.
\bibitem{[6]}
T. Filk, {\it Phys. Lett. B} {\bf 376} (1996) 53.
\bibitem{[8]}
S. Minwalla, M. Van Raamsdonk and N. Seiberg,
{\it J. High Energy Phys.} 00{\bf 02}: 020 (2000).
\bibitem{m}
H. Weyl, {\it Z. Phys.} {\bf 46} (1927) 1.
\bibitem{marm}
O. V. Man'ko, V.I. Man'ko and G. Marmo, {\it Phys. Scr.} {\bf 62} (2000) 446.
\bibitem{marm1}
O. V. Man'ko, V. I. Man'ko and G. Marmo,
{\it J. Phys. A} {\bf 35} (2002) 699.
\bibitem{4}
A. Connes, M. R. Douglas and A. Schwarz, {\it J. High Energy Phys.}
98{\bf 02}: 003 (1998).
\bibitem{12}
M. R. Douglas and C. Hull, {\it J. High Energy Phys.}
98{\bf 02}: 008 (1998).
\bibitem{19}
D. Minic, hep-th/9909022.
\bibitem{20}
A. Jevicki and S. Ramgoolam, {\it J. High Energy Phys.} 99{\bf 04}:
032 (1999); Z. Chang, {\it Eur. Phys. J. C} {\bf 17} (2000) 527;
H. Steinacker, {\it Adv. Theor. Math. Phys.} {\bf 4} (2000) 155.
\bibitem{21}
R. J. Finkelstein, {\it Lett. Math. Phys.} {\bf 38} (1996) 53.
\bibitem{22}
L. Castellani, {\it Phys. Lett. B} {\bf 292} (1992) 93.
\bibitem{sc}
L. Brink and J. H. Schwarz, {\it Phys. Lett. B} {\bf 100} (1981) 310.
\bibitem{FL}
S. Ferrara and M. A. Lled\'o, {\it J. High Energy Phys.}
00{\bf 05}: 008 (2000).
\bibitem{KL}
P. Kosianski, J. Lukierski and P. Maalanka,
Quantum deformations of space-time SUSY
and noncommutative superfield theory, hep-th/0011053.
\bibitem{KP}
D. Klemm, S. Penati and L. Tamassia,
{\it Class. Quant. Grav.} {\bf 20} (2003) 2905.
\bibitem{gord}
A. Nowicki, E. Sorace and M. Tarlini,
{\it Phys. Lett. B} {\bf 302} (1993) 419.
\bibitem{efim1}
V. A. Alebastrov and G. V. Efimov,
{\it Commun. Math. Phys.} {\bf 31} (1973) 1.
\bibitem{star}
J. M. Gracia Bondia, F. Lizzi, G. Marmo and P. Vitale,
{\it J. High Energy Phys.} 02{\bf 04}: 026 (2002).
\bibitem{sup}
A. F. Ferrari et al.,
{\it Phys. Rev. D} {\bf 74} (2006) 125016;
A. F. Ferrari, H. O. Girotti, M. Gomes, A. Yu.
Petrov, A. A. Ribeiro and A. J. da Silva,
{\it Phys. Lett. B} {\bf 577} (2003) 83.
\bibitem{mukhi}
S. Mukhi, {\it Phys. Rev. D} {\bf 20} (1979) 1839.
\bibitem{kovalenko}
S. G. Kovalenko, in {\it Dubna 1992, Weak and Electromagnetic Interactions
in Nuclei}, 505--514.
\bibitem{prokhorov}
L. V. Prokhorov, {\it JETP} {\bf 45} (1963) 791.
\bibitem{joglekar}
S. D. Joglekar, hep-th/0601006; Hai-Jun Wang, quant-ph/0512162.
\bibitem{rs}
B. Mukhopadhyaya, S. Sen, S. Sen and S. SenGupta,
{\it Phys. Rev. D} {\bf 70} (2004) 066009.
\end{thebibliography}
\end{document}